\documentclass{iopart}

\usepackage{graphicx}
\usepackage{amssymb}

\newcommand{\boxeq}[2]{\fbox{$\begin{array}{c}\textrm{#1}\\\textrm{ #2}\end{array}$}}

\begin{document}

\title{Universality of cauliflower-like fronts: from nanoscale thin films to macroscopic plants}

\author{Mario Castro}
\address{Grupo de Din\'amica No-Lineal and Grupo Interdisciplinar de Sistemas Complejos (GISC), Escuela T\'ecnica Superior de Ingenier\'{\i}a (ICAI), Universidad Pontificia Comillas, E-28015, Madrid, Spain}
\author{Rodolfo Cuerno}
\address{Departamento de Matem\'aticas and GISC,
Universidad Carlos III de Madrid, E-28911 Legan\'es, Spain
}
\author{Matteo Nicoli}
\address{Physique de la Mati\`ere Condens\'ee, \'Ecole Polytechnique - CNRS, 91128 Palaiseau, France}
\author{Luis V\'azquez}
\address{Instituto
de Ciencia de Materiales de Madrid, CSIC, Cantoblanco, E-28049 Madrid,
Spain}
\author{Josephus G.\ Buijnsters}
\address{Department of Metallurgy and Materials Engineering, Katholieke Universiteit Leuven, B-3001 Leuven, Belgium}

\begin{abstract}
Chemical vapor deposition (CVD) is a widely used technique to grow solid materials with accurate control of layer thickness and composition. Under 
mass-transport-limited conditions, the surface of thin films thus produced
grows in an unstable fashion, developing a typical motif that resembles the familiar surface of a cauliflower plant. Through experiments on CVD production of amorphous hydrogenated carbon films leading to cauliflower-like fronts, we provide a quantitative assessment of a continuum description of CVD interface growth. As a result, we identify non-locality, non-conservation, and randomness as the main general mechanisms controlling the formation of these ubiquitous shapes. We also show that the surfaces of actual cauliflower plants and combustion fronts obey the same scaling laws, proving the validity of the theory over seven orders of magnitude in length scales. Thus, a theoretical justification is provided, that had remained elusive thus far, for the remarkable similarity between the textures of surfaces found for systems that differ widely in physical nature and typical scales.


\end{abstract}

\maketitle

\section{Introduction}
\label{}

Chemical vapor deposition (CVD) is a technique that is extensively used to grow
films whose surfaces have controlled smoothness or composition \cite{CVDbook}. Part of the
generalized use of CVD to produce coatings or thin films is due to the fact
that it can be used with almost all elements and with many compounds. Basically, CVD involves
the film growth of a solid out from the aggregation of species that appear as the result of the reaction
or decomposition of volatile precursors within a chamber. Chemical reactions
occur in the vicinity of or at the surface of the solid, by-products being removed
when needed. Here, we are interested in CVD as a technique that is capable to grow a surface under
far from equilibrium conditions, yielding unstable rough surfaces that resemble the morphology of a
familiar cauliflower plant~\cite{messier1985}. We will refer to these surfaces as cauliflower-like fronts.

Interestingly, not only growing thin films display this appealing
cauliflower texture, but also many other natural patterns do. In general,
these shapes, although easily recognizable, are not regular but present
some self-similar or hierarchical structure within a characteristic sea of
randomness. In this sense, cauliflower-like fronts rank among the most
fascinating natural forms, in view of their simplicity and considering their diversity in
origins and scales: they can be observed across length scales that range from tens
of nanometers (surfaces of amorphous thin films~\cite{messier1985}) up to
hundreds of microns (turbulent combustion fronts~\cite{dyke1982}) and tens of
centimeters (the familiar cauliflower plants). However, these morphologies
being originated under non-equilibrium conditions, there is a lack of a general theoretical
framework that can account for such a diversity and ubiquity.

Another feature that makes CVD attractive as a benchmark to understand
surface growth far from equilibrium is the possibility to formulate a physically motivated theory for
interface dynamics, which incorporates the essential mechanisms that drive the process when this
production technique is employed~\cite{bales1990}. However, to our knowledge, a detailed comparison between such theory and the mentioned (fractal) cauliflower-like fronts is still lacking. This is remarkable in view of the wide interest that fractal geometry~\cite{mandelbrot1982} has raised in the past, having been recognized to encode the morphological features of self-similar systems, namely those whose structure looks the same with independence of the scale of observation. Actually, many of the best-known fractals ---as, for instance, computational models of biological morphogenesis~\cite{barnsley1993}--- are geometrical structures constructed deterministically by iteration of a simple initial motif. However, this qualitative knowledge is not entirely satisfactory because, as mentioned, cauliflower-like structures are not exactly regular but, rather, appear random to the eye.

In this work we provide a detailed comparison between experimental surfaces of thin amorphous hydrogenated carbon films grown by CVD and predictions from a physical model derived from first principles. Excellent agreement is obtained both for (qualitative) morphological as well as for (quantitative) statistical analysis. This allows us to identify the main features of cauliflower-like fronts, as well as the essential general mechanisms that lead to their occurrence, thus accounting for their ubiquity in natural systems across several orders of magnitude. These conclusions are reached after further morphological analysis of actual cauliflower plants and combustion fronts for which typical scales are macroscopic, rather than submicrometric as in our CVD experiments. The interface evolution equation we consider is thus postulated as a universal description for non-local interface growth under appropriate conditions.

\section{Model}


In Refs.\ \cite{cuerno2001,nicoli2008,nicoli2009b} a generic system of equations for CVD surface growth was presented. Those equations contain the main mechanisms involved in CVD growth: diffusion in the vapor phase, reaction, and attachment via surface kinetics, generalizing the classic description of the process (see \cite{bales1990} and references therein) to account for fluctuation effects in aggregation events and diffusive fluxes. Performing a standard linear and weakly nonlinear analysis, one arrives at a closed equation for the height of the film surface, $h({\bf r},t)$, at time $t$, where ${\bf r}$ is the position above a reference plane. Actually, the equation is more easily expressed for the space Fourier
transform of the surface, $h_{\bf q}(t)\equiv {\mathcal F} [h({\bf x},t)]$ and, using $q=|\mathbf{q}|$, reads
\begin{equation}
\partial_t h_{\bf q}= \left(Vq - D d_0 q^3\right) \, h_{\bf q}+\frac{V}{2}{\mathcal F}\{(\nabla h)^2\}_{\bf q}+\eta_{\bf q}.
\label{mainCVD}
\end{equation}
Here\footnote{See Table II for a glossary and summary of the main variables and parameters used in this work.}, $V$ is the average velocity of the interface, $D$ is the diffusion coefficient in the vapor phase, $d_0$ is the capillarity length of the surface, and the Gaussian white noise term $\eta_{\mathbf{q}}$ contains the information about the underlying microscopic fluctuations \cite{cuerno2001,nicoli2008,nicoli2009b}, having zero mean and correlations given by
\begin{equation}
\langle \eta_{\bf q}(t) \eta_{\bf q'}(t') \rangle = D_n (2\pi)^2 \delta({\bf q}+{\bf q'}) \, \delta(t-t') .
\label{main_noise}
\end{equation}
Despite the apparent simplicity of equation (\ref{mainCVD}), emphasis must be done in its real space representation
in order to stress its non-local character. Thus, the terms proportional to odd powers
of $q$ correspond to fractional Laplacians acting on the height field,
\begin{equation}
(-\nabla^2)^{(2p-1)/2} h(\mathbf{r}) = c_{2,2p-1} \, {\rm PV} \int_{\mathbb{R}^2}
\frac{h(\mathbf{r})-h(\mathbf{r}')}{|\mathbf{r}-\mathbf{r}'|^{2p+1}}
\, {\rm d}\mathbf{r}' , \;\;\; p=1, 2,
\label{lap}
\end{equation}
where ${\rm PV}$ denotes Cauchy principal value and $c_{2,2p-1}$ are appropriate numerical constants \cite{Samko:2002,kassner2005,Silvestre:2007}. Indeed, the Fourier representation of (\ref{lap}) is given by
\begin{equation}
\mathcal{F}[(-\nabla^2)^{(2p-1)/2} h] = q^{2p-1} h_{\mathbf{q}}, \;\;\; p=1, 2,
\label{lap_F}
\end{equation}
as occurring in Eq.\ (\ref{mainCVD}). Thus, in real space these terms couple height differences with algebraically decaying kernels. Hence, the value of the local growth velocity at a given surface point depends on the values of the height at all other surface points. Physically, this non-local coupling arises from the competition of the different parts of the system over the resources for growth~\cite{MBDLA}. In the case of a solid surface growing out of species that aggregate from a vapor phase, it is induced by the geometrical shadowing of prominent surface features that are more exposed to diffusive fluxes, over more shallow ones~\cite{bales1990}\footnote{Similarly, for combustion of premixed flames, the competition is for the available unburnt fuel \cite{bychkov2000}.}.


\begin{figure}[!tp]
\begin{center}
\includegraphics[width=0.75\textwidth,clip=]{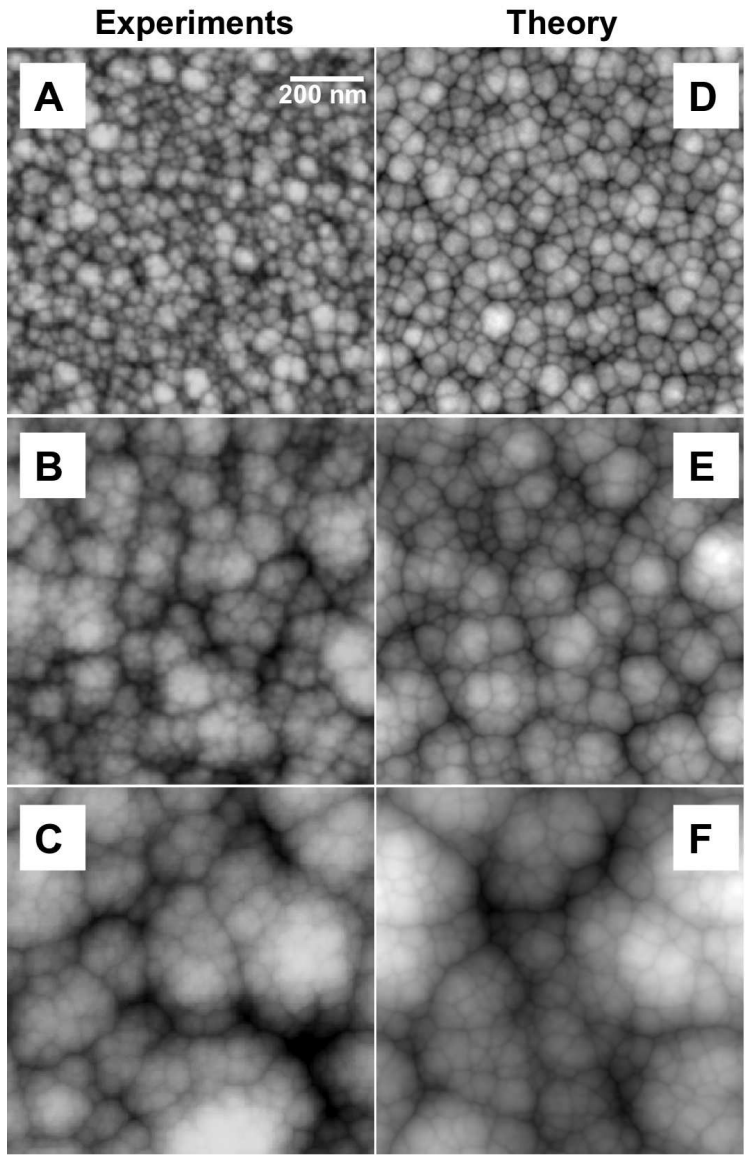}
\end{center}
\caption{Nanocauliflower surface growth. AFM top views ($1\times 1$ $\mu$m$^2$) of CVD experiments at times $t=40$ min (A), $t=2$ hrs (B), and $t=6$ hrs (C). Panels D-F show numerical results from Eq.\ (\ref{mainCVD}) for the same times.}
\label{morphologies}
\end{figure}

\section{Comparison with experiments}

In order to understand the physical implications of Eq.\ (\ref{mainCVD}), we
resort to numerical simulations that circumvent the analytical difficulties
posed by its nonlinearity. Moreover, we have performed growth experiments
in order to show that this equation indeed describes {\em quantitatively}
actual cauliflower-like morphologies in CVD. Specifically, amorphous hydrogenated
carbon (a-C:H) films were grown by electron cyclotron resonance chemical
vapor deposition (ECR-CVD) on silicon substrates in a commercial ECR reactor
(ASTeX, AX4500) in a two-zone vacuum chamber operating with a $2.45$ GHz
microwave source at $208-210$ Watt input power. Gas mixtures of methane/argon
($15$ sccm/$35$ sccm) were applied keeping the operating pressure at
$1.1\times10^{-2}$ Torr. A dc bias of $-50$ V was applied to the silicon
substrates. All samples were grown under these conditions and only the
deposition time was varied. The film surface morphology was characterized {\em
ex-situ} by Atomic Force Microscopy (AFM) with a Nanoscope IIIa equipment
(Veeco) operating in tapping mode with silicon cantilevers.

Top views of the surface morphology are shown in Fig.\ 1, panels A to C. As we see, a ``globular'' structure appears at short times with a characteristic length-scale, that grows in a disorderly fashion with further deposition.
In our ECR-CVD growth system, the main growth species are ions and
radicals. The latter can be distinguished into two main groups, {\em
i.e.}, C$_1$H$_x$ and C$_2$H$_x$ radicals. Within the first subgroup,
C$_1$H$_{2,3}$ radicals have values of the sticking probability $s$ to
become permanently attached to the surface about $s\simeq 10^{-4}-10^{-2}$,
whereas C$_1$H and C$_1$ have a sticking coefficient close to unity~\cite{kim2006red}.
The C$_2$H$_x$ radicals generally have a high sticking coefficient ($s\simeq
0.4-0.8$)~\cite{neyts_influence_2006}. In fact, for a pure methane plasma,
an overall sticking coefficient $s=0.65\pm 0.15$ has been estimated~\cite{hopf1999slp}.
Moreover, when methane is diluted with argon, the impingement of argon ions
generates dangling bonds at the surface leading to an effective increase
of $s$ for the different growth species. Thus, we can assume that the
effective sticking coefficient is close to unity in our system, $s\simeq
1$~\cite{maya_synergistic_2008}. This allows us to determine the surface
kinetics regime at which experiments are operating. Assessment is done
through comparison of the two velocity scales in the system: the mean surface
velocity, $V$, and the mass transfer rate $k_D$, that is related to the
sticking probability $s$ through
\begin{equation}
k_D=DL^{-1}_{\rm mfp}\,\frac{s}{2-s},
\label{mfp}
\end{equation}
with $L_{\rm mfp}$ being the mean free path of molecules in the vapor phase (see \cite{palmer1988} and references therein). From data in Table 1 and Eq.\ (\ref{mfp}), we can estimate $k_D\simeq 0.75$ cm s$^{-1}$, which is considerably higher than $V=2.4\times 10^{-8}$ cm s$^{-1}$. Hence, the system can be assumed to be in the fast kinetics regime for which Eq.\ (\ref{mainCVD}) is expected to hold \cite{cuerno2001,nicoli2008,nicoli2009b}.

\begin{table}
\begin{center}
\begin{tabular}{l c }\hline
Observable& Value\\\hline
Growth velocity ($V$) & $2.4\pm 0.12\times 10^{-8}$ cm s$^{-1}$ ($864\pm 43$ nm/h) \\
Mean substrate temperature ($T$) & 343 K \\
Mean free path ($L_\textrm{mfp}$) & $0.45\pm 0.05$ cm \\
Partial pressure (methane) & $3.75\pm 0.25\times 10^{-3}$ mbar\\
Diffusion coefficient ($D$) & $0.33\pm 0.02$ cm$^2$s$^{-1}$ \\ \hline
\end{tabular}
\end{center}
\caption{ Summary of parameters that can be measured or estimated from our CVD growth experiments.}
\end{table}
\vspace{3mm}

\begin{table}
\begin{center}
\begin{tabular}{|l|l|} \hline
{\bf Name} & {\bf Meaning}\\ \hline
CVD & Chemical Vapor Deposition \\
ECR-CVD & Electron Cyclotron Resonance CVD \\
${\mathcal F}$ & Fourier transform operator \\
$V$ & Average interface growth velocity\\
$D$ & Diffusion coefficients of particles in the gas \\
$d_0$ & Capillarity length \\
$l_c$ & Characteristic lengthscale in the linear regime \\
$D_n$ & Amplitude of the noise fluctuations \\
$C_{2,3,4}$ & Stabilizing coefficients in the general model ({\em e.g.}, $C_3=d_0D$)\\
$h({\bf x},t)$ & Position (height) of the growing interface\\ 
$h_{{\bf q}}(t)$ & Fourier transformation of $h({\bf x},t)$ \\
$\eta({\bf x},t)$ & Noise term accounting for fluctuations\\ 
$\eta_{{\bf q}}(t)$ & Fourier transformation of $\eta({\bf x},t)$ \\
PSD& Power spectral density (also $S({\bf q},t)$) \\
$W(t)$ & Global roughness or width if the interface (standard deviation of the height) \\
$\alpha $ & Roughness exponent \\
$z$ & Dynamic exponent\\
$\beta $ & Growth exponent ($\beta =\alpha /z$)\\  \hline
\end{tabular}
\end{center}
\caption{Summary of the acronyms and main variables used in this work.}
\end{table}

In spite of the previous assessment of parameter values, still many microscopic details of the experimental setup cannot be measured or even estimated from data, mainly due to the limited resolution of the experimental measurements, and also due to the coexistence between species both in the vapor phase and at the very same aggregate surface. To cite a few, the mean atomic volume of aggregating species at the surface, surface tension, or the capillarity length. Unfortunately, some of these are crucial in order to determine the quantitative values of the coefficients in
Eq.\ (\ref{mainCVD}).

Hence, in order to proceed further we must extract additional parameter values from analysis of the morphologies in Fig.\ 1A-C. First, we render Eq.\ (\ref{mainCVD}) non-dimensional by fixing appropriate time, length, and height scales, namely,  
we perform the following change of variables: $x\rightarrow x'\equiv x/x_0$ (so $q\rightarrow q'\equiv x_0q$, $t\rightarrow t'\equiv t/t_0$, $h\rightarrow h'\equiv h/h_0$, so that  Eq.\ (\ref{mainCVD}) reads 
\begin{equation}
\fl
\frac{h_0}{t_0}\partial_{t'}h'_{\bf q'}= \left(\frac{h_0}{x_0}Vq' - D d_0 \frac{h_0}{x_0^3}(q')^3\right) \, h'_{\bf q}+\frac{Vh_0^2}{2x_0^2}{\mathcal F}\{(\nabla' h')^2\}_{\bf q'}+(x_0^{-2}t_0^{-1})^{-1/2}\eta_{\bf q'}.
\label{mainCVD2}
\end{equation}
By properly choosing $x_0$, $t_0$ and $h_0$, as
\begin{equation}
x_0=h_0=\sqrt{\frac{Dd_0}{V}},\quad t_0=\sqrt{\frac{Dd_0}{V^3}},
\end{equation}
we can reduce the latter equation to
(after dropping the primes for convenience)
\begin{equation}
\partial_t h_{\bf q}= \left(q - q^3\right) \, h_{\bf q}+\frac{1}{2}{\mathcal F}\{(\nabla h)^2\}_{\bf q}+\left(\frac{t_0D_n}{h_0^2x_0^2}\right)\tilde\eta_{\bf q},
\label{mainCVD3}
\end{equation}
where $\tilde\eta_{{\bf q}}$ is a white noise term with zero mean and variance $1$. Note how all the information is now contained in the prefactor of the noise term which is, after fixing the lateral size of the simulation  domain (in our case $L=512$, see the discussion below), the only
remaining free parameter to be fitted. We have performed simulations for different values of $D_n$ until we have found the optimal value that
provides the best agreement with the experiments (in our case, $t_0D_n/h_0^2x_0^2=0.21$).

Starting from a flat initial condition, at short times surface slopes are small so that the quadratic nonlinear term in Eq.\ (\ref{mainCVD}) is expected to be negligible. The system will thus evolve according to the linear terms, which has implications on the statistical properties of the morphology. For instance, the computed skewness of the height distribution in Fig.\ 1A is negligible (note how the height distribution in Fig.\ 2A is almost symmetric), which is consistent with a small contribution of the nonlinearity $|\nabla h|^2$, that is the only term breaking the up-down symmetry of the surface. Neglecting this term a characteristic length scale can be identified in Eq.\ (\ref{mainCVD})
that, prior to non-dimensionalization, is given by
\begin{equation}
l_c=2\pi\sqrt{\frac{3Dd_0}{V}} .
\end{equation}
Hence, the ratio between $l_c$ and the lateral system size $L_x$ in numerical simulations of Eq.\ (\ref{mainCVD}) must agree with the ratio between the experimental value $l_c=28$ nm and the experimental AFM window, $L_c=1$ $\mu$m. Thus, we obtain approximately $L_x\simeq 512$ in our dimensionless units (we have rounded this value up to an exact power of $2$ in order to optimize the numerical integration of the equation by means of a pseudo-spectral algorithm). 

The lengthscale $l_c$ is the geometric average of the diffusion length (in
the bulk), $l_D=D/V$) and the capillarity length. Physically, this average
arises from the competition between the scales {\em explored} by the diffusing
particles in the gas and the lengthscales at which they can travel on the
surface until the either aggregate or evaporate. In practice, $l_c$ can be
interpreted as the typical size of the cauliflower-like structures than can be
identified in the surface morphology at short times, see Fig. 1A.

Numerical simulations of Eq.\ (\ref{mainCVD}) using the same scheme as in \cite{nicoli2008} are shown
in Fig.\ 1D-F for the same set of times as for the experimental images that appear in the same figure.
The time evolution of the surface consists of an initial regime controlled by the Mullins-Sekerka linear instability \cite{cross2009} that leads to the appearance of a pattern (cusp) with characteristic length scale $l_c$. In a process that is reminiscent of the stochastic Michelson-Sivashinsky equation that describes
combustion fronts (see~\cite{nicoli2009} and references therein), there is a competition between
cusp coarsening/annihilation, and cusp formation induced by noise, leading to fully non-linear dynamics. As a result, for long enough times unstable growth is stabilized by the quadratic Kardar-Parisi-Zhang (KPZ) nonlinearity~\cite{nicoli2009}, the surface morphology becoming disordered and rough, with height fluctuations that are scale free both in
time and in space~\cite{cuerno2001}. Note the strong resemblance between the theoretical and the experimental morphologies shown in Fig.\ 1.

We have done a more quantitative comparison between Eq.\ (\ref{mainCVD}) and experimental surfaces. In particular, we have determined the distribution of heights (Fig.\ 2), the height power spectral density (PSD, see Figs.\ 3 and 4), $S(q,t)$, defined as 
$S(q,t)=\langle h_{\mathbf{q}}(t)h_{-\mathbf{q}}(t)\rangle$,
where brackets denote average over noise realizations,
and the global width or roughness, $W(t)$ (Fig.\ 3), defined as the standard deviation of the interface height around its mean.

The distribution of heights for short and long times is shown in Fig.\ 2 corresponding to the morphologies shown
in Figs.\ 1A,C,D,F. The distributions are very noisy but, overall, the shape of the curves is comparable for both experiments and theory. As mentioned above,
for short times ($t=40$ minutes) the system is in the linear regime and one can neglect the role of the non-linearity. As a result, all the terms preserve the symmetry $h\rightarrow -h$ and the distributions are symmetric. On the other hand, for long times larger slopes develop as a result of the initial exponential growth and the non-linear term breaks that symmetry. This can be easily seen in Fig.\ 2B.
\begin{figure}[!ht]
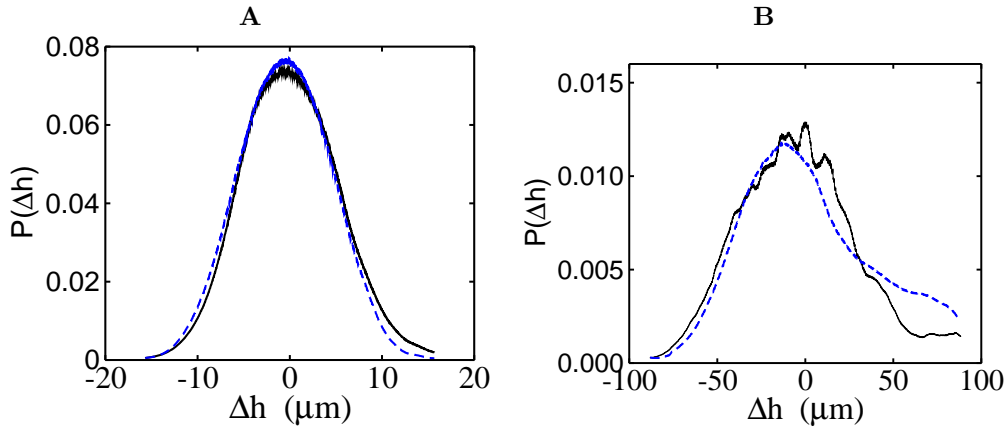

\begin{center}
\begin{tabular}{cc}
{\bf A} & {\bf B}\\
\includegraphics[width=0.49\textwidth,clip=]{heightdistr_ad} & 
\includegraphics[width=0.49\textwidth,clip=]{heightdistr_cf} \\
\end{tabular}
\end{center}
\caption{\label{heightdistr} Normalized distribution of heights, $P(\Delta h)$ for A) Figures 1A (black solid line) and 1D (blue dashed line) corresponding to $t=40$ minutes and B) Figures 1C (black solid line) and 1F  (blue dashed line) for  $t=6$ hours.}
\end{figure}

As discussed in the introduction, cauliflower-like fronts are characterized by scale invariance, which can be quantified with the power spectral density (PSD).
For the time scales of Fig.\ 1, the PSD reflects the scale invariance associated with kinetic roughening (self-affine interfaces), and is expected to behave as $S(q) \sim q^{-(2\alpha+d)}$ \cite{barabasi1995}. Here, $\alpha$ is the so-called roughness exponent and $d=2$ is the substrate dimension. Besides, the roughness grows as a power law of time $W(t)\sim t^{\alpha/z}$, where $z$ is the so-called dynamic exponent that measures the speed at which height correlations spread laterally across the interface \cite{barabasi1995}. It is customary to define a third roughness exponent, $\beta $, that is related to the previous ones through $\beta =\alpha /z$.  For each curve $S(q,t)$ in Fig.\ 3, the small $q$ behavior corresponds to an uncorrelated interface, the crossover to correlated spectra moving to smaller $q$ (larger length scales) as time proceeds. In our case we obtain numerically $\alpha=1.03\pm 0.06$ and $\beta =0.93\pm 0.07$ which are equal, within error bars, to the values $\alpha=\beta =z=1$ predicted by Renormalization Group (RG) calculations on Eq.\ (\ref{mainCVD})~\cite{nicoli2009,nicoli2011}. In order to obtain these exponent values, we have computed the roughness
using the relation
\begin{equation}
W^2(t)=\int S({\mathbf{q},t)\,d\mathbf{q}}=\int 2\pi qS(q,t)dq,
\label{width_def}
\end{equation}
where $S(q,t)$ is the radially averaged PSD (note that the argument here is $q=|{\bf q}|$). Hereafter we will refer to this radially averaged function simply as 2D PSD.
\begin{figure}[!ht]
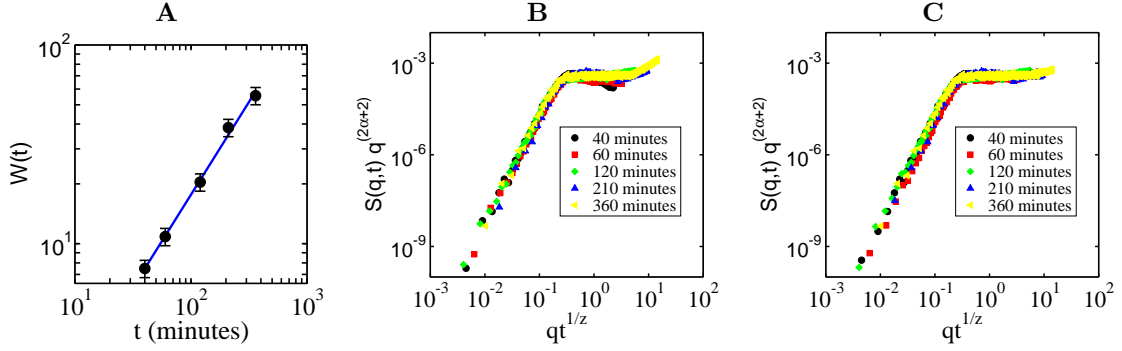

\begin{center}
\begin{tabular}{ccc}
{\bf A} &{\bf B} & {\bf C}\\
\includegraphics[width=0.32\textwidth,clip=]{width2}&
\includegraphics[width=0.37\textwidth,clip=]{colapso} &
\includegraphics[width=0.37\textwidth,clip=]{colapsoth} \\
\end{tabular}
\end{center}
\caption{\label{widthcollapse} A) Experimental global width or roughness, $W(t)$ (solid circles). The blue solid line has slope $\beta=0.93\pm 0.07$. B) Collapse of the radially averaged power spectral density for times from $t=40$ minutes to $6$ h (see legend) obtained for $\alpha =1.03$ and $z=\alpha /\beta $, with $\beta$ as obtained from $W(t)$. C) Same as panel B) but for the theoretical model.}
\end{figure}

In Fig.\ 3A we show experimental time evolution of the surface roughness. As shown, the value obtained for the growth exponent $\beta =0.93\pm 0.07$ is close to the theoretical prediction $\beta =1$. A customary method to determine the roughness exponents is by means of the collapse of the PSDs at different times, by properly scaling $S({\bf q},t)$ and $q$ as shown in Figs.\ 3B (experiments) and Fig.\ 3C (theory).

The quantitative agreement that we obtain between the experimental and theoretical PSD functions is not limited to the values of the scaling exponents. Remarkably, it actually extends to the behavior of the full functions along the dynamics of the system, as can be seen in Fig. 4, where we compare theoretical and experimental PSDs for short and long times. Such type of agreement goes much beyond what is usually expected on the context of universal properties  \cite{barabasi1995}.

\begin{figure}[!tp]
\begin{center}
\includegraphics[width=0.95\textwidth,clip=]{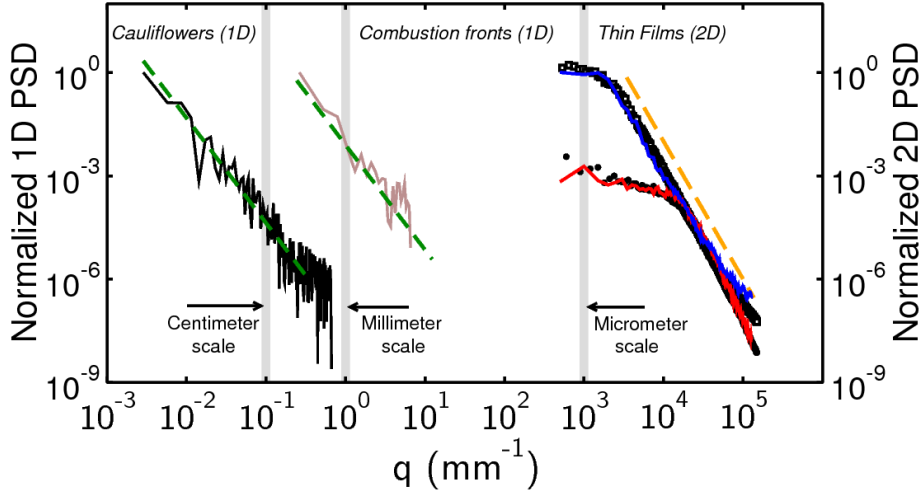}
\end{center}
\caption{Scaling universality of cauliflower surface growth. Normalized PSD functions are shown to compare systems spanning several orders of magnitude in size. Relevant scales are indicated with arrows. The label {``Thin Films''} identifies the comparison between the experimental and theoretical (angular average of the) PSD for thin films grown by CVD. Circles: experiment $t = 40$ min. Solid red line: theory $t = 40$ min. Squares: experiment $t = 6$ h. Dashed blue line: theory t = 6 h. The orange dashed straight line is a guide to the eye with slope $-4$ as predicted by Eq.\ (\ref{mainCVD}). The solid black line under the label ``Cauliflowers'' has been obtained after averaging results obtained for 10 cauliflower slices ({\em Brassica Oleracea}, from two different specimens). The solid brown line under {``Combustion fronts''} corresponds to the PSD of an experimental combustion profile~\cite{law_cellular_2005}. The green dashed lines have slopes $-3$. This value differs from the $-4$ obtained for the Thin Films case because we are computing the PSD of one-dimensional slices in the cases of the cauliflowers and combustion fronts.}
\label{psd}
\end{figure}

\section{Universality of cauliflower-like fronts}
\label{}

Thus far we have been using the term {\em cauliflower} in a loose way. In
order to justify this usage it is convenient to identify the main universal
phenomena and morphological properties that occur in the emergence of cauliflower-like structures
in other contexts, and compare them with those assessed in the previous section. For
instance, in the case of the familiar plants one can postulate (i) an
interaction among the {\em branches} that sustain the external surface.
This interplay would induce {\em competitive growth} among different
plant features; moreover (ii) {\em mass is non conserved} and (iii) {\em
fluctuations} are intrinsic to the biological underlying processes taking
place both at the level of the cell metabolism and in the interaction with the
environment. (iv) An extra stabilizing ingredient is necessary in order to
guarantee the dynamical stability of the ensuing surface (whose specific form,
as we argue below, does not change the statistical properties of that surface).
In line with the occurrence of universality in the properties of rough
surfaces evolving far from equilibrium \cite{barabasi1995}, for appropriate
cases the argument can be reversed. Thus, when comparing two different
systems, the same statistical properties of the surface are a manifestation of
the same governing general principles, in spite of the fact that the detailed physical
mechanisms controlling the dynamics of, e.g., aggregating species in CVD and plant cells, are quite
different indeed.

This property is a generalization of what happens in the proximity of a critical point within the framework of critical phenomena in equilibrium systems. Thus, while the microscopic details are different, the character of the interactions (in our case imposed by non-local competition or non-conservation) dictates the dynamics.
For instance, in the case of the cauliflowers, non-locality is caused by branch competition. In the case of CVD, it stems from the fact that the diffusing particles 
access with a higher probability the most exposed parts of the surface.
 Finally, in the case of combustion, the parts of the front which lie behind the average have less access to oxygen and other combustion species.

The basic ingredients expected for cauliflower-like surface growth, (i) through (iii) above, should reach a non-trivial balance resulting generically into a morphology that, albeit disordered, presents a self-similar, hierarchical structure. Moreover, one expects a typical characteristic length-scale to arise at the finest observation scale, due to the competition between stable and unstable growth mechanisms, as generically occurs in pattern forming systems \cite{cross2009}.

Schematically,
\begin{eqnarray}
\boxeq{Variation of}{height} &=&\boxeq{Non-locality}{(competition)}+\boxeq{Stabilizing mechanism}{(short scale)}\nonumber\\
&&+\boxeq{Non-conservation}{(non-linearity)}+\boxeq{Fluctuations}{(noise)}
\label{schematic}
\end{eqnarray}

If we were to formulate a general interface equation that incorporates these mechanisms, 
actually arguments exist that can impose restrictions on its possible mathematical form. First, we assume that such an equation is weakly non-linear, in the sense that it is a polynomial in small powers of the height and its derivatives. This is a standard simplification in the study both of scale-invariant spatially extended systems \cite{kardar2007}, and of pattern forming systems \cite{cross2009} in the long wavelength limit close to instability threshold, although see \cite{misbah2010} and below. Next, self-similarity requires system statistics to remain unchanged under amplification of the lateral length scale of the sample by a factor $b$, while rescaling the surface height by the same factor,
$\langle h(b{\bf r})\rangle=\langle bh({\bf r})\rangle$,
where $\langle\cdot\rangle$ means {\em in a statistical sense}. For instance, in the case of cauliflower plants, one would expect the exposed surface to arise as an envelope for an underlying branched structure. This branching structure provides {\em volume-filling} mechanisms that guarantee efficient distribution of energy and nutrients.

Moving further, the dominant {\em linear} term in the sought-for equation of motion can be inferred with large generality through dimensional analysis. Thus, assuming there is a single velocity scale, $V$, involved in the surface growth, the (linear) rate of amplification, $\omega_{\lambda}$, of a fluctuation can be related with the typical length-scale of the perturbation, $\lambda$, as
\begin{equation}
\omega_{\lambda}\lambda V^{-1}=\textrm{dimensionless constant}\Rightarrow \omega_{q}\sim Vq.
\label{disp-rel}
\end{equation}
This expression is traditionally referred to as a dispersion relation, and is
often written in terms of the wave number $q\equiv |{\bf q}|=2\pi \lambda^{-1}$.
Actually, Eq.\ (\ref{disp-rel}) ensues for the celebrated Diffusion-Limited Aggregation model (DLA) that is the paradigm of fractal growth \cite{DLA}, and contains the signatures of unstable growth and non-local branch competition. In addition, further stabilizing mechanisms contribute to Eq.\ (\ref{disp-rel}) as higher powers of $q$. Physical examples of non-local growth include solidification from a melt~\cite{langer1980}, flame fronts~\cite{michelson1977}, stratified fluids~\cite{ostrovsky1983}, thin film evolution due to crystalline stress~\cite{shchukin1999}, viscous fluid fingering~\cite{kessler1988}, growing biomorphs~\cite{biomorphs}, or geological structures~\cite{misbah2004}, to cite a few.

The final ingredient for the height equation of motion is {\em non-linearity}. The natural choice is the KPZ term $(V/2)(\nabla h)^{2}$, that has been argued to be generically present in the continuum description of surfaces that grow irreversibly in the absence of conservation laws~\cite{kardar1986} and has been recently assessed to a high degree of accuracy in one-dimensional experiments~\cite{takeuchi2011}. Likewise, the simplest expression of fluctuations is through a random (uncorrelated) function of space and time like a Gaussian distributed white noise $\eta(\mathbf{r},t)$.

Combining all these general ingredients together, we can write down the evolution equation for the {\em local} surface
velocity, that in view of Eq.\ (\ref{disp-rel}) takes a particularly simple form when written for the Fourier
modes of the height and noise fields, namely,
\begin{equation}
\partial_t h_{\bf q}= \left[Vq+\sum_{j=2}^4 C_j q^j\right] \, h_{\bf q}+\frac{V}{2}{\mathcal F}\{(\nabla h)^2\}_{\bf q}+\eta_{\bf q},
\label{main}
\end{equation}
where $C_j$ are negative constants that depend on the specific stabilizing physical conditions. Note that Eq.\ (\ref{mainCVD}) simply corresponds to the particular case of Eq.\ (\ref{main}) in which $C_2=C_4=0$. An important result is that the same values of the scaling exponents $\alpha=z=1$ occur for any stabilizing linear mechanism of the form $C_j q^j$ with $j\geq 2$, as indicated by RG analysis~\cite{nicoli2009}. Consequently, we are confident that this scaling behavior can be also identified in other systems for which the stabilizing term may have different non-zero contributions $C_{j}$. For instance, combustion fronts \cite{michelson1977} and stratified fluids~\cite{ostrovsky1983} correspond to $C_2\neq 0$ and $C_{3}=0$, and one again finds $\alpha =z=1$. 

In practical terms, all the terms in the main equation cooperate to produce those self-similar structures. Thus, if we replace the term $Vq$ for, for instance, $vq^2$ then one would have obtained the celebrated Kuramoto-Sivashinsky equation. Or, if one suppresses the nonlinear term ($\frac{V}{2}{\mathcal F}[\nabla h)^2]$ then the numerical integration will explode as the surface roughness would increase exponentially without control.

As an example, in Fig.\ 4 we also show the power spectral density of the profile of an expanding spherical flame (the experimental profile was taken from Ref.\ \cite{law_cellular_2005}),
showing good agreement with the predictions of Eq.\ (\ref{main}). Additionally, we have performed statistical measurements of the outermost surfaces of cauliflower plants. Specifically, we have computed the power spectral density of the interface profiles of several slices from specimens of {\em Brassica Oleracea}. The roughness exponent $\alpha$ can be determined by analyzing the power spectral density of these slices. In Fig.\ 4 we also show the averaged PSD over 10 different slices. The slope for small values of $q$ leads to the value $\alpha = 1.02 \pm 0.05$ that, within experimental uncertainty~\cite{kim2005}, is the same as the one provided by our continuum description. Overall, Fig.\ 4 proves
the validity of the generalized Eq.\ (\ref{main}) to describe cauliflower-like fronts across seven orders of magnitude in length scales, finally justifying the use of the term ``cauliflower'' as applied to morphologies that can differ quite strongly from the familiar plants. The difference between these and CVD films or combustion fronts is that, for the former, we have not been able to obtain a dynamical characterization of the plant morphology (we have only characterized it at a fixed, long time), as opposed to the latter in which the full dynamical equation Eq.\ (\ref{main}) has been derived from first principles and has been also experimentally validated, both in the context of CVD (this work) and for combustion fronts, see~\cite{searby}.

\begin{figure}[!tp]
\begin{center}
\includegraphics[width=0.95\textwidth,clip=]{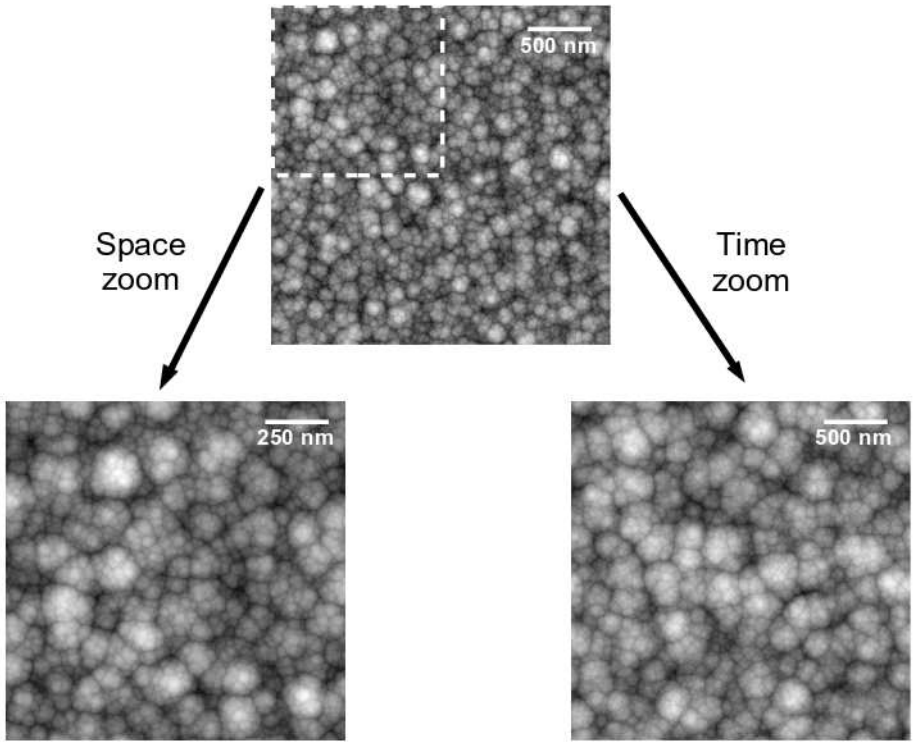}
\end{center}
\caption{Fractality in space and time reflects the inability to distinguish between image zooms and time shifts. This is the hallmark of cauliflower-like surface growth as described by Eq.\ (\ref{main}).
Top: Numerical simulation for $t_1=2$ h. Bottom right: Numerical simulation for $t_2=4$ h. Bottom left: Spatial amplification of the top panel by a factor $t_2/t_1=2$. The zoomed area is indicated in the top panel with a white dashed square.}
\label{zooms}
\end{figure}

To better understand the significance of the unit values found for both critical exponents $\alpha$ and $z$, note that
for kinetically rough, {\em self-affine} surfaces, $W/L\sim L^{\alpha-1}$ for a sufficiently large lateral observation scale, $L$ \cite{barabasi1995}. Precisely for $\alpha=1$ the system is not merely self-affine but becomes, rather, {\em self-similar}, its geometrical features remaining statistically invariant in the macroscopic limit $L\rightarrow \infty$. Note moreover that in this case the average growth velocity for Eq.\ (\ref{main}) is only due to the nonlinear term and becomes scale-independent precisely for $\alpha=1$, showing the self-consistency of our initial assumption on a single velocity scale. Also the fact that $z=1$ reflects another peculiar fact about the fractality of the system: the system is self-similar also in time. Namely, correlations travel ballistically across the surface so that time behaves as space under rescaling. Hence, if we observe the system at two different times, we cannot distinguish the second one from an isotropic spatial zoom performed in the earlier one, as illustrated in Fig.\ \ref{zooms}. This corresponds to the
intuition that by mere visual inspection it is hard to distinguish between the whole cauliflower plant and a piece of it, and between young and small florets. Therefore, in spite of the lack of dynamical information about cauliflower plant growth, the fact that self-similarity in time constraints the value of $z$ to be unity, and the confirmation of $\alpha\simeq 1$ from Fig.\ 3, both give us confidence to suggest that the growth of cauliflower plants obeys the same general principles as CVD growth.

\section{Conclusions}$\quad$
To conclude, we have shown that Eq.\ (\ref{mainCVD}) provides an accurate description of unstable thin film growth by CVD, agreeing with experiments both qualitatively as well as quantitatively. To our knowledge, this is the first time in which such a quantitative agreement between theory and experiment is achieved, that goes beyond values of critical exponents, reaching the full dynamical behavior of observables such as the power spectral density. Actually, the moving boundary problem leading to Eq.\ (\ref{mainCVD}) has quite a generic form that is relevant to a number of processes in which
transport is diffusive, such as solidification from a melt or electrochemical deposition (see references e.g.\ in \cite{nicoli2009}), so that similar quantitative descriptions of different growth systems by Eq.\ (\ref{mainCVD})
can be foreseen. Beyond that, we have also seen that a similar scaling behavior characterized by scaling exponents $\alpha=z=1$ can be moreover described by Eq.\ (\ref{main}) that applies to other systems that differ in the (linear) relaxation mechanisms.

We would like to emphasize some important points concerning the implications of Eq.\ (\ref{main}). It is intriguing that the geometrical properties of cauliflower-like structures are at the boundary between disorder and fractality, between self-affinity and self-similarity. Thus, the values $\alpha=z=1$ of the scaling exponents induce an interface which is disordered at all scales, while allowing at the same time for the identification of a ``typical'' texture or motif. RG calculations and numerical simulations \cite{nicoli2009,nicoli2011} both indicate the robustness of these exponents values, suggesting the universality of Eq.\ (\ref{main}) as a description of a large class of non-equilibrium systems. Note, however, that interfaces developed under the same general physical principles as elucidated here, but for which the evolution equation is strongly, rather than weakly, nonlinear, may feature different morphological properties from the present cauliflower type. Examples are known in the dynamics of thin \cite{kassner2002} and epitaxial films, and are reviewed in \cite{misbah2010}.

One of the reasons why fractals are so popular is the promise that, knowing their generating rules, we can infer the {\em character of the} underlying physical or biological mechanisms.  Hence whether the interactions are non-local vs local, non-conserved vs conserved, self-similar vs self-affine, \ldots  dictates the form of the mathematical equations.
In contrast to ``algorithmic'' descriptions of fractals, the virtue of our continuum dynamical formulation is that it allows us to extract which are the most relevant mechanisms~\cite{kadanoff1986} whose interplay gives rise to these appealing structures, namely, non-locality, non-conservation, and noise. Among all the possible mathematical forms of non-locality, self-similarity enforces $\alpha =1$. This conclusion is expected to guide the inference of the relevant mechanisms at play in specific physical or biological systems where cauliflower-like structures are identified. Moreover, it is remarkable that such a simple equation as Eq.\ (\ref{main}) can be able to capture this non-trivial dynamics, to the extent that, by means of pseudo-spectral numerical integration, the system is capable of efficiently producing realistic patterns that resemble turbulent flame fronts or the texture of cauliflower plants.

From a more general point of view, our theory also brings up the long-standing question as to why natural evolution favors self-similar structures. The so-called allometric scaling relations~\cite{west1997} explain (and predict) the {\em branching} structure of living bodies. The central idea behind these theories is that biological time scales are limited by the rates at which energy can be spread to the places where it is exchanged with the tissues. Thus, the {\em space-filling} structure~\cite{murray2007} required to supply matter and energy to a living system can be accounted for. Focusing on more specific systems (cauliflower plants, etc.), albeit with a large degree of universality, our work suggests the self-similar features that the ``canopy'' atop such branched structures may have.


\ack

This work has been partially supported through Grants No.\ FIS2009-12964-C05-01,-03, and -04 (MICINN, Spain). 
J. G. B. would like to thank the Executive Research Agency of the European Union for funding under the Marie Curie IEF grant number 272448.

\end{document}